\documentclass{jnmp01b}

\usepackage{amsmath}

\numberwithin{equation}{section}

\allowdisplaybreaks

\newcommand{\p}[1]{(\ref{#1})}

\makeatletter
\DeclareRobustCommand{\primfrac}[1]{%
  \PackageWarning{amsmath}{%
Foreign command \@backslashchar#1; %
\protect\frac\space or \protect\genfrac\space should be used instead%
  }
  \global\@xp\let\csname#1\@xp\endcsname\csname @@#1\endcsname
  \csname#1\endcsname
}
\makeatother

\begin{document}

\renewcommand{\evenhead}{O. Lechtenfeld and A. Sorin}
\renewcommand{\oddhead}{Real Forms of  
Supersymmetric Toda Chain Hierarchy}


\thispagestyle{empty}

\begin{flushleft}
\footnotesize \sf
Journal of Nonlinear Mathematical Physics \qquad 2000, V.7, N~4,
\pageref{firstpage}--\pageref{lastpage}.
\hfill {\sc Letter}
\end{flushleft}

\vspace{-5mm}

\copyrightnote{2000}{O. Lechtenfeld and A. Sorin}

\Name{Real Forms of the Complex Twisted N=2 Supersymmetric Toda Chain 
Hierarchy in Real N=1 and Twisted N=2 Superspaces}

\label{firstpage}

\Author{O. LECHTENFELD~$^\dag$ and A. SORIN~$^\ddag$}

\Adress{$^\dag$ Institut f\"ur Theoretische Physik, Universit\"at Hannover,\\
~~Appelstra{\ss}e 2, D-30167 Hannover, Germany \\
~~lechtenf@itp.uni-hannover.de\\[10pt]
$^\ddag$ Bogoliubov Laboratory of Theoretical Physics, JINR,\\
~~141980 Dubna, Moscow Region, Russia \\
~~sorin@thsun1.jinr.ru}

\Date{Received March 1, 2000; Revised March 29, 2000; 
Accepted May 22, 2000}

\begin{abstract}
\noindent
Three nonequivalent real forms of the complex twisted N=2 supersymmetric
Toda chain hierarchy (solv-int/9907021) in real N=1 superspace are presented. 
It is demonstrated that they possess a global twisted N=2 supersymmetry. 
We discuss a new superfield basis in which the supersymmetry transformations 
are local. Furthermore, a representation of this hierarchy is given in terms 
of two twisted chiral N=2 superfields.   
The relations to the s-Toda hierarchy by H. Aratyn, E. Nissimov and 
S. Pacheva (solv-int/9801021) as well as to the modified and derivative NLS 
hierarchies are established.
\end{abstract}


\section{Introduction}

Recently an $N{=}(1|1)$ supersymmetric generalization of the 
two-dimensional Darboux transformation was proposed in \cite{ls}
in terms of $N=(1|1)$ superfields, and an infinite class of bosonic and
fermionic solutions of its symmetry equation was constructed in \cite{ls}
and \cite{ols}, respectively. These solutions generate bosonic and
fermionic flows of the complex $N=(1|1)$ supersymmetric Toda lattice
hierarchy\footnote{A wide class of the complex Toda
lattices connected with Lie superalgebras was first
introduced in the pioneering papers \cite{olshanetsky,leites,andreev}
(see also recent papers \cite{evans} and references therein).}
 which actually possesses a more rich symmetry, namely 
complex $N=(2|2)$ supersymmetry. Its one-dimensional reduction possessing
complex $N=4$ supersymmetry ---the complex $N=4$ Toda chain
hierarchy--- was discussed in \cite{dgs}. There, the Lax
pair representations of the bosonic and fermionic flows, the corresponding
local and nonlocal Hamiltonians, finite and infinite discrete symmetries,
the first two Hamiltonian structures and the recursion operator were
constructed. Furthermore, its nonequivalent real forms in real $N=2$
superspace were analyzed in \cite{ds}, where the relation to the complex 
$N=4$ supersymmetric KdV hierarchy \cite{di} was established. 

Consecutively, the reduction of the complex $N=4$ supersymmetric Toda chain
hierarchy from complex $N=2$ superspace to complex $N=1$ superspace was
analyzed in \cite{ols1}, where also its Lax-pair and Hamiltonian
descriptions were developed in detail. Here, we call this
reduction the {\it complex twisted N=2 supersymmetric Toda chain
hierarchy}, due to the common symmetry properties of its three different
real forms which will be discussed in what follows (see the paragraph after
eq. \p{conjn23}). The main goals of the present letter are firstly to
analyze real forms of this hierarchy in real $N=1$
superspace with one even and one odd real coordinate, secondly to derive a
manifest twisted $N=2$ supersymmetric representation of its simplest
non-trivial even flows in twisted $N=2$ superspace, and thirdly to clarify
its relations (if any) with other
known hierarchies (s-Toda \cite{anp1}, modified NLS and derivative NLS
hierarchies).  

Let us start with a short summary of the results that we shall need
concerning the complex twisted $N=2$ supersymmetric Toda chain hierarchy
(see \cite{ols1,dgs,ls,ols} for more details).

The complex twisted $N=2$ supersymmetric Toda chain hierarchy in 
complex $N=1$ superspace comprises an infinite set of even and odd
flows for two complex even $N=1$ superfields
$u(z,\theta)$ and $v(z,\theta)$, where
$z$ and $\theta$ are complex even and odd
coordinates, respectively. The flows are generated by complex
even and odd evolution derivatives
$\{{\textstyle{\partial\over\partial t_k}}, ~U_k\}$ and $\{ D_k,~Q_k\}$
($k \in \mathbb{N}$), respectively, with the following length
dimensions:
\begin{equation}
[{\textstyle{\partial\over\partial t_k}}]=[U_k]=-k, \quad
[D_k]=[Q_k]=-k+\frac{1}{2},
\label{dimtimes}
\end{equation}
which are derived by the reduction of the supersymmetric KP
hierarchy in $N=1$ superspace \cite{maninradul}, 
characterized by the Lax operator 
\begin{equation}
L=Q+v D^{-1}u.
\label{lax1}
\end{equation}
$D$ and $Q$ are the odd covariant derivative
and the supersymmetry generator, respectively,
\begin{equation}
D\equiv \frac{\partial}{\partial {\theta}} +
{\theta} {\partial}, \quad
Q\equiv \frac{\partial}{\partial {\theta}} -
{\theta} {\partial}.
\label{algDQ0}
\end{equation}%
\resetfootnoterule%
They form the  algebra\footnote{We explicitly present only non-zero
brackets in this letter.}
\begin{equation}
\{ D,D\} = +2{\partial}, \quad \{ Q,Q\} = -2{\partial}.
\label{alg00}
\end{equation}
The first few of these flows are:
\begin{gather}
{\textstyle{\partial\over\partial t_0}}
\left(\begin{array}{cc} v\\ u \end{array}\right) =
\left(\begin{array}{cc} +v\\ -u \end{array}\right), \quad
{\textstyle{\partial\over\partial t_1}}
\left(\begin{array}{cc} v\\ u \end{array}\right) =
{\partial}\left(\begin{array}{cc} v\\ u \end{array}\right),
\label{eqs1}
\\[1ex]
\begin{split}
&{\textstyle{\partial\over\partial t_2}} v =
+v~'' -  2uv(DQv)+(DQv^2u)+v^2(DQu) -2v(uv)^2,
\\
&{\textstyle{\partial\over\partial t_2}} u =
-u~'' -  2uv(DQu)+(DQu^2v)+u^2(DQv) +2u(uv)^2,
\end{split}
\label{eqs}
\\[1ex]
\begin{split}
{\textstyle{\partial\over\partial t_3}} v ={}&
v~''' +3(Dv)~'(Quv)-3(Qv)~'(Duv)
+3v~'(Du)(Qv) \\
&-3v~'(Qu)(Dv)+6vv~'(DQu)-6(uv)^2v~', \\
{\textstyle{\partial\over\partial t_3}} u ={}&
u~''' +3(Qu)~'(Duv)-3(Du)~'(Quv)
+3u~'(Qv)(Du) \\
&-3u~'(Dv)(Qu)
+6uu~'(QDv)-6(uv)^2u~',
\end{split}
\label{flow3}
\\[1ex]
\begin{split}
&
D_1 v= -Dv+ 2vQ^{-1}(uv),
\quad D_1 u= -Du- 2uQ^{-1}(uv), \\
&
Q_1 v= -Qv- 2vD^{-1}(uv),
\quad Q_1 u= -Qu+ 2uD^{-1}(uv), 
\end{split}
\label{ff2-}
\\[1ex]
U_0\left(\begin{array}{cc} v\\ u \end{array}\right) =
{\theta}D \left(\begin{array}{cc} v\\ u \end{array}\right).
\label{qqq1}
\end{gather}
Throughout this letter, we shall use the notation $u'=\partial 
u={\partial\over\partial z}u$. 
Using the explicit expressions of the flows \p{eqs1}--\p{qqq1}, one
can calculate their algebra which has the following nonzero brackets:
\begin{gather}
\Bigl\{D_k\,,\,D_l\Bigr\}=
-2\;\frac{{\partial}}{{\partial t_{k+l-1}}}, \quad
\Bigl\{Q_k\,,\,Q_l\Bigr\}=
+2\;\frac{{\partial}}{{\partial t_{k+l-1}}},
\label{alg1}
\\[1ex]
\Bigl[U_k\,,\,D_l\Bigr]=Q_{k+l}, \quad
~\Bigl[U_k\,,\,Q_l\Bigr]=D_{k+l}.
\label{algqqbar}
\end{gather}
This algebra produces an affinization of the algebra of global
complex $N=2$ supersymmetry, together with an affinization of its
$gl(1,\mathbb{C})$ automorphisms. 
It is the algebra of symmetries of the nonlinear even flows
\p{eqs}--\p{flow3}. The generators may be realized in the superspace
$\{t_k,\theta_k,\rho_k, h_k \}$,
\begin{equation}
\begin{split}
& D_k= \frac{\partial}{\partial \theta_k}- \sum^{\infty}_{l=1}\theta_l
\frac{\partial}{{\partial t_{k+l-1}}},\quad
Q_k=\frac{\partial}{\partial {\rho}_k}+
\sum^{\infty}_{l=1}{\rho}_l
\frac{\partial}{{\partial t_{k+l-1}}}, \\
& 
U_k=\frac{\partial}{\partial h_k}-
\sum^{\infty}_{l=1}({\theta}_l
\frac{\partial}{{\partial {\rho}_{k+l}}}+{\rho}_l
\frac{\partial}{{\partial {\theta}_{k+l}}}),
\end{split}
\label{covder}
\end{equation}
where $t_k, h_k$ ($\theta_k,\rho_k$) are bosonic (fermionic) abelian
evolution times with length dimensions
\begin{equation}
[t_k]=[h_k]=k, \quad [\theta_k] =[\rho_k]=k-\frac{1}{2}
\label{dim}
\end{equation}
which are in one-to-one correspondence with the length dimensions
\p{dimtimes} of the corresponding evolution derivatives.

The flows $\{{\textstyle{\partial\over\partial t_k}},~D_k,~Q_k\}$ 
can be derived from the flows 
$\{{\textstyle{\partial\over\partial t_k}},~D^{+}_k,~D^{-}_k\}$ 
of the complex $N=4$ Toda chain hierarchy \cite{dgs} by the reduction
constraint 
\begin{equation}
{\theta}^{+}=i{\theta}^{-}\equiv {\theta} 
\label{ccccc}
\end{equation}
which leads to the correspondence $D_+\equiv D$ and $D_-\equiv iQ$ with the 
fermionic derivatives of the present paper, where $i$ is the imaginary unit 
and ${\theta}^{\pm}$ are the Grassmann coordinates of the $N=2$ superspace
in \cite{dgs}.

\section{Real forms of the complex twisted N=2 Toda chain
hierarchy} 

It is well known that different real forms derived from the
same complex integrable hierarchy are nonequivalent in general.
Keeping this in mind it seems important to find as many
different real forms of the complex twisted $N=2$ Toda chain hierarchy as
possible.

With this aim let us discuss various nonequivalent complex
conjugations of the superfields $u(z,\theta)$ and
$v(z,\theta)$, of the superspace coordinates $\{z,~\theta\}$,
and of the evolution derivatives
$\{{\textstyle{\partial\over\partial t_k}},~U_k,~D_k, ~Q_k\}$
which should be consistent with the flows \p{eqs1}--\p{qqq1}.
We restrict our considerations to the case
when $iz$ and $\theta$ are coordinates of real $N=1$ superspace
which satisfy the following standard complex conjugation properties:
\begin{equation}
(iz,{{\theta}})^{*}=(iz,{\theta}).
\label{conj}
\end{equation}
We will also use the standard convention regarding complex conjugation of
products involving odd operators and functions (see, e.g., the books
\cite{ggrs}). In particular, if $\mathbb{D}$ is some even differential
operator acting on a superfield $F$, we define the complex conjugate of 
$\mathbb{D}$ by $(\mathbb{D}F)^*=\mathbb{D}^*F^*$. Then, in the
case under consideration one can derive, for example, the following
relations
\begin{equation}
\begin{split}
&{\partial}^*=-{\partial}, \quad
{\epsilon}^{*}={\epsilon}, \quad
{\varepsilon}^{*}={\varepsilon}, \quad
({\epsilon}{\varepsilon})^{*}=-{\epsilon}{\varepsilon},\\
&({\epsilon}D)^{*}={\epsilon}D, \quad
({\varepsilon}Q)^{*}={\varepsilon}Q, \quad (DQ)^{*} = - DQ
\end{split}
\label{conjrel}
\end{equation}
which we use in what follows. Here, ${\epsilon}$ and ${\varepsilon}$ 
are constant odd real parameters.

Let us remark that, although most of the flows of the
complex twisted $N=2$ supersymmetric Toda chain hierarchy can be derived by
reduction \p{ccccc}, its real forms in $N=1$ superspace \p{conj} cannot be
derived in this way from the real forms of the complex $N=4$ Toda chain
hierarchy in the real $N=2$ superspace 
\begin{equation}
(iz,{{\theta}}^{\pm})^{*}=(iz,{\theta}^{\pm})
\label{conjn4}
\end{equation}
found in \cite{ds}. This conflict arises because the constraint \p{ccccc} is
inconsistent with the reality properties \p{conjn4} of the $N=2$
superspace.   

We would like to underline that the flows \p{eqs1}--\p{qqq1} form a
particular realization of the algebra \p{alg1}--\p{algqqbar} in terms of
the $N=1$ superfields $u(z,\theta)$ and $v(z,\theta)$. Although
the classification of real forms of affine and conformal
superalgebras was given in a series of classical papers
\cite{new1,new2} (see also interesting paper \cite{new3} for
recent discussions and references therein) we cannot
obtain the complex conjugations of the target space superfields
$\{u(z,\theta), v(z,\theta)\}$ using only this base. It is
a rather different, non-trivial task to construct the
corresponding complex conjugations of various realizations
of a superalgebra which are relevant in the context of
integrable hierarchies. Moreover, different complex
conjugations of a given (super)algebra realization may
correspond to the same real form of the (super)algebra, while
some of its other existing real forms may not be reproducible
on the base of a given particular realization. In what follows
we will demonstrate that this is exactly the case for the
realization under consideration. We shall see that complex conjugations
of the target space superfields $\{u(z,\theta), v(z,\theta)\}$
correspond to the twisted real $N=2$ supersymmetry.

Direct verification shows that the flows \p{eqs1}--\p{qqq1}
admit the following three nonequivalent complex conjugations
(meaning that it is not possible to relate them via obvious symmetries):
\begin{gather}
\begin{split}
&(v,u)^{*}= (v,-u), \quad
(iz,{{\theta}})^{*}=(iz,{\theta}), \\
&(t_p,U_p,{\epsilon}_p D_p,
{\varepsilon}_pQ_p)^{*}=(-1)^{p}(t_p,U_p,
-{\epsilon}_pD_p, -{\varepsilon}_pQ_p), 
\end{split}
\label{conj1}
\\[1ex]
\begin{split}
& (v,u)^{\bullet}= (u,v), \quad
(iz,{{\theta}})^{\bullet}=(iz,{\theta}), \\
&(t_p,U_p,{\epsilon}_p D_p,
{\varepsilon}_pQ_p)^{\bullet}=(-t_p,U_p,
{\epsilon}_p D_p,{\varepsilon}_pQ_p), 
\end{split}
\label{conj2}
\\[1ex]
\begin{split}
& (v,~u)^{\star}=(~-u(QD\ln u+uv),~ \frac{1}{u}~),
\quad (iz,{\theta})^{\star}=(iz,{\theta}),\\
& (t_p,U_p,{\epsilon}_p
D_p,{\varepsilon}_pQ_p)^{\star}=(-t_p,U_p,
-{\epsilon}_p D_p, -{\varepsilon}_pQ_p),
\end{split}
\label{conj3}
\end{gather}
where ${\epsilon}_p$ and ${\varepsilon}_p$
are constant odd real parameters. We would like to underline that
the complex conjugations of the evolution derivatives (the second
lines of eqs. \p{conj1}--\p{conj3} ) are defined and fixed
completely by the explicit expressions \p{eqs1}--\p{qqq1} for the flows. 
These complex conjugations extract three different real
forms of the complex integrable hierarchy we started with,
while all the real forms of the flows algebra \p{alg1}--\p{algqqbar}
correspond to the same algebra of a twisted global real
$N=2$ supersymmetry.  
This last fact becomes obvious if one uses the $N=2$ basis 
of the algebra with the generators
\begin{equation}
{\cal D}_1\equiv \frac{1}{\sqrt{2}}(Q_1+D_1), \quad
{\overline {\cal D}}_{1}\equiv \frac{1}{\sqrt{2}} (Q_1-D_1).
\label{su(2)}
\end{equation}
Then, the nonzero algebra brackets \p{alg1}--\p{algqqbar} and the
complex conjugation rules \p{conj1}--\p{conj3} are the 
standard ones for the twisted $N=2$ supersymmetry
algebra together with its non--compact $o(1,1)$ automorphism,
\begin{gather}
\Bigl\{{\cal D}_1\,,\,{\overline {\cal D}}_1\Bigr\}=
2 \;\frac{{\partial}}{{\partial t_{1}}}, \quad
\Bigl[U_0\,,\,{\cal D}_1\Bigr]=+{\cal D}_1, \quad \
~~\Bigl[U_0\,,\,{\overline {\cal D}}_1\Bigr]=-{\overline {\cal D}}_1,
\label{algnn2}
\\[1ex]
({\textstyle{\partial\over\partial t_1}},
U_0,{\gamma}_1{\cal D}_1, {\overline {\gamma}}_1
{\overline {\cal D}}_1)^{*}=
(-{\textstyle{\partial\over\partial t_1}},U_0,
+{\gamma}_1{\cal D}_1,
+{\overline {\gamma}}_1 {\overline {\cal D}}_1), \quad
({\gamma}_1, {\overline {\gamma}}_1)^{*}=
({\gamma}_1, {\overline {\gamma}}_1),
\label{conjn21}
\\[1ex]
({\textstyle{\partial\over\partial t_1}},
U_0,{\gamma}_1{\cal D}_1, {\overline {\gamma}}_1
{\overline {\cal D}}_1)^{\bullet}=
(-{\textstyle{\partial\over\partial t_1}},U_0,
+{\gamma}_1{\cal D}_1,
+{\overline {\gamma}}_1 {\overline {\cal D}}_1), \quad
({\gamma}_1, {\overline {\gamma}}_1)^{\bullet}=
({\gamma}_1, {\overline {\gamma}}_1),
\label{conjn22}
\\[1ex]
({\textstyle{\partial\over\partial t_1}},
U_0,{\gamma}_1{\cal D}_1, {\overline {\gamma}}_1
{\overline {\cal D}}_1)^{\star}=
(-{\textstyle{\partial\over\partial t_1}},U_0,
-{\gamma}_1{\cal D}_1,
-{\overline {\gamma}}_1 {\overline {\cal D}}_1), \quad
({\gamma}_1, {\overline {\gamma}}_1)^{\star}=
({\gamma}_1, {\overline {\gamma}}_1),
\label{conjn23}
\end{gather}
where ${\gamma}_1, {\overline {\gamma}}_1$ are constant odd
real parameters. Therefore, we conclude that the complex twisted $N=2$
supersymmetric Toda chain hierarchy with the complex conjugations 
\p{conj1}--\p{conj3} possesses twisted real $N=2$ supersymmetry.
For this reason we like to call it the ``twisted $N=2$
supersymmetric Toda chain hierarchy'' (for the supersymmetric Toda chain
hierarchy possessing untwisted $N=2$ supersymmetry see \cite{bs1} 
and references therein).

Let us remark that a combination of the two involutions
(\ref{conj3}) and (\ref{conj2}) generates the infinite-dimensional group
of discrete Darboux transformations \cite{ols1}
\begin{equation}
\begin{split}
&(v,~u)^{\star \bullet}=(~v(QD\ln v-uv),~ \frac{1}{v}~), \quad
(z,{{\theta}})^{\star \bullet}=(z,{\theta}),\\
& 
(t_p,U_p,D_p,Q_p)^{\star \bullet }=(t_p, U_p,-D_p,-Q_p).
\end{split}
\label{discrsymm}
\end{equation}
This way of deriving discrete symmetries was proposed
in \cite{s} and applied to the construction of discrete symmetry
transformations of the $N=2$ supersymmetric GNLS hierarchies.

To close this section let us stress once more that we cannot claim
to have exhausted {\it all} complex conjugations of the twisted $N=2$ 
Toda chain hierarchy by the three examples of complex conjugations 
(eqs. \p{conj1}--\p{conj3}) we have constructed. Finding complex
conjugations for affine (super)algebras themselves is a problem 
solved by the classification of \cite{new1} but rather different from
constructing complex conjugations for different {\it realizations} of
affine (super)algebras. To our knowledge, no algorithm yet exists for
solving this rather complicated second problem. Thus, classifying {\it all}
complex conjugations is out of the scope of the present letter.
Rather, we have constructed these examples in order to use them merely as
tools to generate the important discrete symmetries \p{discrsymm} as well
as to construct a convenient superfield basis and a manifest twisted $N=2$
superfield representation (see Sections 3 and 4), with the aim to clarify
the relationships of the hierarchy under consideration to other physical
hierarchies discussed in the literature (see Section~5).

\section{A KdV-like basis with locally realized
supersymmetries.}

The third complex conjugation \p{conj3} looks rather complicated
when compared to the first two ones \p{conj1}--\p{conj2}.
However, it drastically simplifies in another superfield basis defined as
\begin{equation}
J\equiv uv + QD\ln u, \quad {\overline J}\equiv -uv,
\label{basis}
\end{equation}
where $J \equiv J(z,\theta)$ and $ {\overline J}\equiv 
{\overline J}(z,\theta)$ ($[J]=[{\overline J}]=-1$) are 
unconstrained even $N=1$ superfields.
In this basis the complex conjugations \p{conj1}--\p{conj3} 
and the discrete Darboux transformations \p{discrsymm} are given by
\begin{gather}
(J,~{\overline J})^{*}=-(J,~{\overline J}),
\label{conj1j}
\\[1ex]
(J,~{\overline J})^{\bullet}=(~J - QD\ln {\overline J}, ~{\overline J}~),
\label{conj2j}
\\[1ex]
(J,~{\overline J})^{\star}=({\overline J},~J),
\label{conj3j}
\\[1ex]
(J,~{\overline J})^{\star \bullet}=(~{\overline J},
~J - QD\ln {\overline J}~),
\label{discrsymm1}
\end{gather}
and the equations \p{eqs}--\p{qqq1} become simpler as well,
\begin{gather}
\begin{split}
&{\textstyle{\partial\over\partial t_2}} J =
(-J~' +2J D^{-1}Q{\overline J}-J^2)~', \\
&{\textstyle{\partial\over\partial t_2}}{\overline J} =
(+{\overline J}~'  + 2{\overline J}D^{-1}QJ-{\overline J}^2)~',
\end{split}
\label{eqs2j}
\\[1ex]
\begin{split}
&{\textstyle{\partial\over\partial t_3}} J =
3~\Bigl[ ~\frac{1}{3}J~'' +J~J~'-J~'D^{-1}Q{\overline J} -
2J^2D^{-1}Q{\overline J}-JD^{-1}Q{\overline J}^2 
+\frac{1}{3}J^3 ~\Bigr]~', \\
&{\textstyle{\partial\over\partial t_3}} {\overline J} =
3~\Bigl[ ~\frac{1}{3}{\overline J}~'' -{\overline J}~{\overline J}~'+
{\overline J}~'D^{-1}QJ-2{\overline J}^2D^{-1}QJ
-{\overline J}D^{-1}QJ^2+\frac{1}{3}{\overline J}^3 ~\Bigr]~',
\end{split}
\label{eqs2jt3}
\end{gather}
and then
\begin{gather}
D_1
\left(\begin{array}{cc} J\\ {\overline J} \end{array}\right) =
D\left(\begin{array}{cc} +J\\ -{\overline J} \end{array}\right),
\quad Q_1 \left(\begin{array}{cc} J \\ {\overline J}  \end{array}\right) =
Q\left(\begin{array}{cc} + J \\ - {\overline J}  \end{array}\right),
\label{supersflowsj}
\\[1ex]
U_0 \left(\begin{array}{cc} J \\ {\overline J}  \end{array}\right) =
{\theta}D \left(\begin{array}{cc} J \\ {\overline J}  \end{array}\right).
\label{qqqj1}
\end{gather}
Notice that the supersymmetry and $o(1,1)$ 
transformations \p{supersflowsj}--\p{qqqj1} of the superfields
$J$, $\bar J$ are local functions of the superfields. 
The evolution equations \p{eqs2j}--\p{eqs2jt3} are also local
because the operator $D^{-1}Q$ is a purely differential one,
$D^{-1}Q\equiv [\theta,D]$.

\section{A manifest twisted N=2 supersymmetric representation}

The existence of a basis with locally and linearly realized twisted $N=2$
supersymmetric flows \p{supersflowsj} would give evidence in favour
of a possible description of the hierarchy  
in terms of twisted $N=2$ superfields. It turns out that this is indeed
the case. In order to show this, let us introduce a twisted $N=2$ 
superspace with even coordinate $z$ and two odd real coordinates $\eta$
and $\overline \eta$  (${\eta}^{*}={\eta},~ {\overline {\eta}}^{*}=
{\overline {\eta}}$), as well as odd covariant derivatives ${\cal D}$ and
${\overline {\cal D}}$ via
\begin{equation}
{\cal D} \equiv \frac{\partial}{\partial {\eta}}+
{\overline \eta} {\partial}, \quad
{\overline {\cal D}}\equiv \frac{\partial}
{\partial {\overline \eta}}+{\eta} {\partial}, \quad
\Bigl\{{\cal D}\,,\,{\overline {\cal D}}\Bigr\}=2{\partial}, \quad
{\cal D}^2={\overline {\cal D}}^2=0
\label{algnn4}
\end{equation}
together with twisted $N=2$ supersymmetry generators 
${\cal Q}$ and ${\overline {\cal Q}}$
\begin{equation}
{\cal Q} \equiv \frac{\partial}{\partial {\eta}}-
{\overline \eta} {\partial}, \quad
{\overline {\cal Q}}\equiv \frac{\partial}
{\partial {\overline \eta}}-{\eta} {\partial}, \quad
\Bigl\{{\cal Q}\,,\,{\overline {\cal Q}}\Bigr\}=-2{\partial}, \quad
{\cal Q}^2={\overline {\cal Q}}^2=0.
\label{algnngen4}
\end{equation}
In this space, we consider two chiral even twisted
$N=2$ superfields ${\{\cal J}(z,\eta,\overline \eta)
~{\overline {\cal J}}(z,\eta,\overline \eta)\}$, which obey 
\begin{equation}
{\cal D}{\cal J}=0, \quad {\cal D} {\overline {\cal J}}= 0
\label{N=2constr}
\end{equation}
and are related to the $N=1$ superfields 
$\{J(z,\theta),~ {\overline J}(z,\theta)\}$ \p{basis}.
More concretely,
their independent components are related to those of 
$J$ and ${\overline J}$ as follows,
\begin{equation}
\begin{split}
&{\cal J}|_{\eta=\overline \eta=0} = J|_{\theta=0}, \quad
{\overline {\cal D}} ~{\cal J}|_{\eta=\overline \eta=0} = +D J|_{\theta=0},
\\ 
&{\overline {\cal J}}|_{\eta=\overline \eta=0}
={\overline J}|_{\theta=0}, \quad
\overline {\cal D}~{\overline {\cal J}}|_{\eta=\overline \eta=0} =
-D{\overline J}|_{\theta=0}. 
\end{split}
\label{N=4n2rel}
\end{equation}
Then, in terms of these superfields the equations
\p{eqs2j}--\p{eqs2jt3} become
\begin{gather}
\begin{split}
&{\textstyle{\partial\over\partial t_2}} {\cal J} =
(-{\cal J}~' -2{\cal J} {\overline {\cal J}}-{\cal J}^2)~', \\
&{\textstyle{\partial\over\partial t_2}}{\overline {\cal J}} =
(+{\overline {\cal J}}~'  - 2{\cal J}{\overline {\cal J}}-
{\overline {\cal J}}^2)~',
\end{split}
\label{eqs2jN=4}
\\[1ex]
\begin{split}
{\textstyle{\partial\over\partial t_3}} {\cal J} =
3~\Bigl( ~\frac{1}{3}{\cal J}~'' +{\cal J}~{\cal J}~'+
{\overline {\cal J}} {\cal J}~'+ 2{\cal J}^2{\overline {\cal J}}+
{\cal J}{\overline {\cal J}}^2 
+\frac{1}{3}{\cal J}^3 ~\Bigr)~', \\
{\textstyle{\partial\over\partial t_3}} {\overline {\cal J}} =
3~\Bigl( ~\frac{1}{3}{\overline {\cal J}}~'' -
{\overline {\cal J}}~{\overline {\cal J}}~'-
{\cal J}{\overline {\cal J}}~'+2{\overline {\cal J}}^2{\cal J}+
{\overline {\cal J}}~{\cal J}^2+\frac{1}{3}{\overline {\cal J}}^3~\Bigr)~',
\end{split}
\label{eqs2jN=4t3}
\end{gather}
and it is obvious that they and the chirality constraints \p{N=2constr} are
manifestly invariant with respect to the transformations generated by the
twisted $N=2$ supersymmetry generators ${\cal Q}$ and 
${\overline {\cal Q}}$ \p{algnngen4}.

Let us also present a manifestly twisted $N=2$ supersymmetric form of  
the complex conjugations \p{conj1j}--\p{conj3j} and the discrete
Darboux transformations \p{discrsymm1} in terms of the superfields 
${\cal J}(z,\eta,\overline \eta)$ and 
${\overline {\cal J}}(z,\eta,\overline \eta)$ \p{N=4n2rel}: 
\begin{gather}
({\cal J},~{\overline {\cal J}})^{*}=-({\cal J},~{\overline {\cal J}}),
\label{conj1jj}
\\[1ex]
({\cal J},~{\overline {\cal J}})^{\bullet}=
(~{\cal J}-{\partial}\ln{\overline {\cal J}}, ~{\overline {\cal J}}~),
\label{conj2jj}
\\[1ex]
({\cal J},~{\overline {\cal J}})^{\star}=({\overline {\cal J}},~{\cal J}),
\label{conj3jj}
\\[1ex]
({\cal J},~{\overline {\cal J}})^{\star \bullet}=
(~{\overline {\cal J}},~{\cal J} -{\partial}\ln {\overline {\cal J}}~),
\label{discrsymm2}
\end{gather}
modulo the standard automorphism which changes the sign of all
Grassmann odd objects.

\section{Relation with the s-Toda, modified NLS and derivative NLS
hierarchies} 

It is well known that there are often hidden relationships
between a priori unrelated hierarchies. Some examples are the $N=2$ NLS
and $N=2$ ${\alpha}=4$ KdV \cite{kst}, the ``quasi'' $N=4$ KdV and 
$N=2$ ${\alpha}= -2$ Boussinesq \cite{dgi}, the $N=2$ (1,1)-GNLS
and $N=4$ KdV \cite{s,bs}, the $N=4$ Toda and $N=4$ KdV \cite{ds}. These
relationships may lead to a deeper understanding of the hierarchies. They
may help to obtain a more complete description and to derive solutions. 
 
The absence of odd derivatives in the equations  
\p{eqs2jN=4}--\p{eqs2jN=4t3}, starting off the twisted $N=2$
supersymmetric Toda chain hierarchy, gives additional evidence in favour
of a hidden relationship with some bosonic hierarchy. It
turns out that such a relationship indeed exists.
Let us search it first at the level of the Darboux transformations
\p{discrsymm2}, then in the second flow equation \p{eqs2jN=4}.

For this purpose, we introduce new $N=1$ superfields 
$\{\Phi(z,\theta), {\Psi}(z,\theta)\}$ via 
\begin{equation}
\begin{split}
&{\cal J}|_{\eta=\overline \eta=0} \equiv 
(\Phi \Psi + \partial \ln \Psi)|_{\theta=0}, \quad
{\overline {\cal D}}~{\cal J}|_{\eta=\overline \eta=0} \equiv 
D(\Phi \Psi + \partial \ln \Psi)|_{\theta=0}, \\
& {\overline {\cal J}}|_{\eta=\overline \eta=0} 
\equiv -(\Phi \Psi)|_{\theta=0}, \quad \quad \quad \quad 
{\overline {\cal D}}~ {\overline {\cal J}}|_{\eta=\overline \eta=0} 
\equiv -D(\Phi \Psi)|_{\theta=0}.
\end{split}
\label{basisn1}
\end{equation}
The Darboux transformations \p{discrsymm2}, expressed in terms of those 
new superfields, exactly reproduce the Darboux-Backlund (s-Toda)
transformations 
\begin{equation}
(\Phi,~\Psi)^{\star \bullet}=(~\Phi(\partial \ln \Phi- \Phi \Psi),~
\frac{1}{\Phi}~) 
\label{discrsymm3}
\end{equation}
proposed in \cite{anp1} in the context of the reduction
of the supersymmetric KP hierarchy in $N=1$ superspace characterized 
by the Lax operator 
\begin{equation}
L=D-2(D^{-1}\Phi \Psi)+\Phi D^{-1}\Psi.
\label{lax2}
\end{equation}
For completeness, we also present the corresponding second flow equations,
\begin{equation}
{\textstyle{\partial\over\partial t_2}} \Phi =
+\Phi~'' -  2\Phi^2\Psi~' -2(\Phi\Psi)^2\Phi ,
\quad {\textstyle{\partial\over\partial t_2}} \Psi =
-\Psi~'' -  2\Psi^2\Phi~' +2(\Phi\Psi)^2\Psi,
\label{eqsar}
\end{equation}
which also follow from \cite{anp1}.
Therefore, we are led to the conclusion that the two integrable hierarchies 
related to the reductions \p{lax1} and \p{lax2} are {\it equivalent}.
It would be interesting to establish a relationship (if any) between
these two hierarchies in the more general case where $v,u$ and $\Phi,\Psi$
entering the corresponding Lax operators \p{lax1} and \p{lax2} 
are rectangular (super)matrix-valued superfields \cite{ols1}, but
this rather complicated question is outside the scope of the present letter. 
In general, these two families of $N=2$ supersymmetric hierarchies
correspond to a non-trivial supersymmetrization\footnote{By trivial
supersymmetrization of bosonic equations we mean just replacing functions
by superfunctions. In this case the resulting equations are supersymmetric,
but they do not contain fermionic derivatives at all.} of bosonic
hierarchies, except for the simplest case we consider here. 
Indeed, a simple inspection shows that the
equations \p{eqs2jN=4}--\p{eqs2jN=4t3} do not contain fermionic
derivatives and belong to the hierarchy which is the trivial $N=2$ 
supersymmetrization of the bosonic modified NLS or derivative NLS
hierarchy. This last fact becomes obvious if one introduces yet a new
superfield basis $\{b(z,\eta,\overline \eta), 
{\overline b}(z,\eta,\overline \eta)\}$ through 
\begin{equation}
{\cal J}\equiv (\ln {\overline b})~', \quad {\overline {\cal J}}\equiv 
-b{\overline b}, \quad
{\cal D}b={\cal D} {\overline b}= 0,
\label{mnlsbasis}
\end{equation}
in which the second flow (\ref{eqs2jN=4}) and the Darboux transformations
\p{discrsymm2} become
\begin{gather}
{\textstyle{\partial\over\partial t_2}} b =
+b~'' + 2b{\overline b}b~', \quad
{\textstyle{\partial\over\partial t_2}} {\overline b}
 =-{\overline b}~'' +  2{\overline b}b{\overline b}~',
\label{eqsmnls}
\\
b^{\star \bullet \star \bullet}=b~(\ln b^{\star \bullet})~', \quad 
{\overline b}^{\star \bullet \star \bullet} = \frac{1}{b}, 
\label{discrsymm4}
\end{gather}
respectively, and the equation \p{eqsmnls} reproduces the trivial $N=2$
supersymmetrization of the modified NLS equation \cite{cll}. When passing
to alternative superfields 
$g(z,\eta,\overline \eta)$ and ${\overline g}(z,\eta,\overline \eta)$
defined by the following invertible transformations
\begin{equation}
g =  b~ {\exp (-{\partial^{-1}} (b {\overline b}))}, \quad
{\overline g} = {\overline b}~{ \exp (+{\partial^{-1}}
(b {\overline b}))},
\label{comp2}
\end{equation}
equation \p{eqsmnls} becomes
\begin{equation}
\frac{\partial}{\partial t_2}g = (+g~' + 2g{\overline g}g)~', \quad
\frac{\partial}{\partial t_2}{\overline g} =
(-{\overline g}~' + 2{\overline g}g{\overline g})~' 
\label{comp3}
\end{equation}
and coincides with the derivative NLS equation \cite{kn}. 

Finally, we would like to remark that one can produce the {\it non-trivial}
$N=2$ supersymmetric modified KdV hierarchy by secondary reduction even
though the twisted $N=2$ Toda chain hierarchy is a {\it trivial} $N=2$
supersymmetrization of the modified or derivative NLS hierarchy. 
One of such reductions was described in
\cite{ols1}. In terms of the superfields $J$
and ${\overline J}$ \p{basis}, the reduction constraint is 
\begin{equation}
J+{\overline J}=0,
\label{red1}
\end{equation}
and only half of the flows from the set \p{dimtimes} are consistent
with this reduction, namely
\begin{equation}
\{~{\textstyle{\partial\over\partial t_{2k-1}}}
~U_{2k},~D_{2k},~Q_{2k}~\}
\label{consistfl}
\end{equation}
(for details, see \cite{ols1}). Substituting the constraint \p{red1} into
the third flow equation \p{eqs2jt3} of the reduced hierarchy, this flow 
becomes
\begin{equation}
{\textstyle{\partial\over\partial t_3}} u =
(J~'' +3(QJ)(DJ)-2J^3)~'.
\label{redflow3}
\end{equation}
Now, one can easily recognize that the equation for the bosonic component
reproduces the modified KdV equation and does not contain the fermionic
component at all. Nevertheless, it seems
that the supersymmetrization \p{redflow3} is rather non-trivial, because it 
involves the odd operators $D$ and $Q$ but does not admit  
odd flows having length dimension $[D]=[Q]=-1/2$. Hence, it does not
seem to be possible to avoid a dependence of $D$ and $Q$ in a cleverly chosen 
superfield basis. To close this discussion let us mention that the
possible alternative constraint on the twisted $N=2$ superfields ${\cal J}$
and ${\overline {\cal J}}$, namely
\begin{equation}
{\cal J}+{\overline {\cal J}}=0,
\label{red2}
\end{equation}
leads again to the trivial $N=2$ supersymmetrization of the modified KdV
hierarchy.

\section{Conclusion}

In this letter we have described three distinct real forms of the 
twisted $N=2$ Toda chain hierarchy introduced in \cite{ols1}. It has been
shown that the symmetry algebra of these real forms is the twisted 
$N=2$ supersymmetry algebra. We have introduced a set of $N=1$ superfields.
They enjoy simple conjugation properties and allowed us to
eliminate all nonlocalities in the flows. All flows and complex
conjugation rules have been rewritten directly in twisted $N=2$
superspace. As a byproduct, relationships between the twisted $N=2$ Toda
chain, s-Toda, modified NLS, and derivative NLS hierarchies have been
established. These connections enable us to derive new real forms of the
last three hierarchies, possessing a twisted $N=2$ supersymmetry.

\subsection*{Acknowledgments}

A.S. would like to thank the Institut f\"ur Theoretische Physik,
Universit\"at Hannover for the hospitality during the course of this work.
This work was partially supported by the DFG Grant No. 436 RUS 113/359/0
(R), RFBR-DFG Grant No. 99-02-04022,
the Heisenberg-Landau programme HLP-99-13, PICS Project No. 593, RFBR-CNRS
Grant No. 98-02-22034, RFBR Grant No. 99-02-18417, Nato Grant No. PST.CLG
974874 and INTAS Grant INTAS-96-0538.

\label{lastpage}

\end{document}